\documentclass[12pt]{article}     

\input epsf

\let\a=\alpha   \let\d=\delta
\let\e=\epsilon   
   \let\l=\lambda
\let\m=\mu \let\n=\nu  \def\p{\vec{p}} \def\q{\vec{q}}
 \let\s=\sigma   
   
 \let\S=\Sigma   \let\L=\Lambda
\let\G=\Gamma \let\D=\Delta  

\def\2{{1\over2}} \def\4{{1\over4}} \def\52{{5\over2}} \def\6{\partial }
\def\({\left(} \def\){\right)} \def\<{\langle } \def\>{\rangle }

\def\nn{\nonumber}

\def\beg{\begin{equation}}
\def\begar{\begin{eqnarray}}
\def\ee{\end{equation}}
\def\ea{\end{eqnarray}}

\newcommand{\pref}[1]{(\ref{#1})}                
\newcommand{\plabel}[1]{\label{#1}}              
\newcommand{\pcite}[1]{\cite{#1}}                
\newcommand{\pbib}[1]{\bibitem{#1}}              
\renewcommand{\Im}{\mbox{Im}}

\begin{document}

\begin{titlepage}
\hfill TUW-96-05
 \begin{center}
\Large
 {\bf On the Spectrum of Scalar-Scalar Bound States}\\[1cm]
\normalsize
 W. M\"odritsch$^*$\\[1.2cm]
 Institut f\"ur Theoretische Physik, Techn. Univ. Wien\\
 A-1040 Wien, Wiedner Hauptstra\ss e 8-10\\
~\\

\end{center}
\vspace{1.0cm}
\centerline{Abstract}

A new, exactly solvable, Barbieri-Remiddi like equation for
bound states of two scalar constituents interacting with 
massless vector particles is presented, both for stable and unstable
particles. With the help of this equation the bound state spectrum is 
calculated to $O(\a^4)$ for a SU(N) nonabelian gauge theory. 
The result for the abelian case reproduces the known result from the
Foldy-Wouthuysen calculation. It is shown how different graphs as in the
fermionic theory contribute to the spectrum to this order. 
Furthermore the bound state correction to the decay width for a 
weakly decaying system is calculated.
This result is equal to its fermionic counterpart. Thus the theorem
on bound state corrections for weakly decaying particles, formulated 
previously for fermions only, has been extended to the scalar theory.

\vspace{\fill}

\_\hrulefill \hspace*{8cm} \\
revised version, Wien, Dec. 1996\\
\\
\\
$^*$ e-mail: wmoedrit@tph16.tuwien.ac.at
\end{titlepage}

\section{Introduction}

While the discussion of fermionic bound states has a long history 
\pcite{Bethe}, much less attention has been paid to  
the similar problem with scalar constituents. Only the ladder 
approximation with scalar interaction is a well known example and has 
already been discussed in the 50-s and 60-s \pcite{Wick} in the framework 
of the Bethe-Salpeter equation. 
Indeed, to this day the only known fundamental matter fields are 
fermionic. But in supersymmetric theories
for each fermion two scalar partners are required.
Since some of them, probably stop or sbottom, could have masses within 
the reach of the next generation of $e^+ e^-$ accelerators, even
the observation of bound states  of those particles seems
possible. These objects and systems built of scalar composite particles in atomic
physics underline the need of an equally clear and transparent
approach as the one developed for the fermionic case \pcite{BR}. A recent 
attempt in this direction \pcite{Owen} splits the boson propagator in a 
particle and anti-particle propagator in order to be able to treat them like fermions.
The spectrum is then obtained by constructing the Hamiltonian via a Foldy -
Wouthuysen transformation and a perturbation theory \`a la Salpeter.
This approach does not show significat advantage over the pure 
Foldy-Wouthuysen approach \pcite{Bjorken} and suffers also from the drawback 
that it will break down in higher orders due to the appearance of higher 
powers in the spatial momentum $\p$. The Coulomb field appears in this 
formalism as an external field which makes this formalism not very 
reliable-looking.
All these drawbacks can be circumvented by developing an 
exactly solvable zero order equation and subsequently using a systematic
perturbation theory. 

To the best of our knowledge there exists no attempt in the literature to construct a
solvable zero order equation for the BS equation containing two
charged scalars interacting via a vector field.
This goal will be achieved in section 2. 

In section 3 we will review 
briefly the BS perturbation theory and use it to calculate the spectrum
of bound states for scalar particles with equal mass, both 
for the abelian and nonabelian case to $O(\a ^4)$.  This will be of 
importance if the stop has a narrow width. If the width becomes comparable
to the level splittings this considerations can be understood as a 
determination of the scalar-antiscalar potential. 

The decay width is also subject of the second application we present 
in section 4. We calculate the bound state correction to the decay width 
$\G$ of system of scalar constituents to $O(\a^2 \G)$.

Finally section 5 is devoted to the conclusions and to the discussion of
our results.

\section{A bound state equation for scalar particles} 

\subsection{Stable particles}

As starting
point we present here an exactly solvable equation
for stable scalar particles which interact via a vector field. 

We start from the BS equation for a bound state wave function $\chi$
\beg \plabel{allg}
\chi_{ij}^{BS}(p;P) = -i S_{ii'}(\frac{P}{2}+p)S_{j'j}(-\frac{P}{2}+p)
\int \frac{d^4p'}{(2\pi)^4}         K_{i'j',i''j''}(P,p,p')
\chi_{i''j''}^{BS}(p';P),
\ee
where $S$ is the exact scalar propagator,
and $K$ is the sum of all two scalar irreducible graphs. Both are 
normalized to be Feynman amplitudes. Furthermore, we
have introduced relative momenta $p$ and $p'$, a total momentum
$P=p_1-p_2$, and we choose the center of mass (CM) frame where
$P=(P_0,\vec{0})=(2m+E,\vec{0})$. 

As a first approximation to eq. \pref{allg} we would like to use beside 
the free relativistic
scalar propagators the kernel due to the Coulomb interaction
\beg \plabel{kcs}
K_C(p,p') =  4\pi \a \frac{(P_0+p_0+p_0')(P_0-p_0-p_0')}{(\p-\p\,')^2}.
\ee
For a nonablelian theory with gauge group $SU(N)$ we use 
\beg \plabel{alph}
  \a = \frac{N^2-1}{2N} \frac{g^2}{4 \pi}.
\ee
In this case $\chi$ has to be a singlett in order that $K_C$ represents
an attractive force.
The kernel \pref{kcs} has the drawback that it is $p_0$ dependent and the
exact solution of eq. \pref{allg} with \pref{kcs} is not known. However,
in the nonrelativistic regime by the scaling argument \pcite{Kum1}
\begar p_0 &\approx& O(m\a^2), |\p|
\approx O(m\a),  \plabel{scal}\\
P_0 &\approx& 2m-O(m\a^2) \nn
\ea
we can start with an instantaneous
approximation to the kernel since $p_0$ is of $O(\a^2 m)$ in this region
and may be included in the corrections afterwards.
Doing this, we can perform the zero component integration on the propagator
( $E_p = \sqrt{m^2+\p\,^2}$)
\begar
&-i& \int \frac{dp_0}{2  \pi} \frac{1}{[(\frac{P_0}{2}+p_0)^2-E_p^2+i\e]
 [(-\frac{P_0}{2}+p_0)^2-E_p^2+i\e]} =  \nn \\
&=& \frac{1}{2 E_p P_0}\left[\frac{1}{2E_p-P_0}-\frac{1}{2E_p+P_0} \right] =
 \frac{1}{E_p (4  E_p^2-P_0^2)}  \plabel{sprop}
\ea
and it is quite easy to show that
\beg \plabel{k0s}
K_0(p,p') = 4\pi \a \frac{4 m \sqrt{E_p E_{p'}}}{\q\,^2}
\ee
gives a solvable equation with the normalized solutions
\begar
\chi(p) &=& i \frac{\sqrt{E_p} (P_0^2-4E_p^2)}{\sqrt{2P_0}
          [(\frac{P_0}{2}+p_0)^2-E_p^2+i\e][(-\frac{P_0}{2}+p_0)^2-E_p^2+i\e]}
          \phi(\p) \\
\bar{\chi}(p,\e) &=& -\chi^*(p,-\e)  \plabel{chisbar}
\ea
to the eigenvalues
\beg
 P_0 = M_n^{(0)} = 2m \sqrt{1-\s_n^2}, \qquad \s_n = {\a \over {2 n}}. 
 \plabel{Pn}
\ee
Eq. \pref{chisbar} is dictated by the requirement that $\bar{\chi}$
should acquire the same analytic properties as the underlying field
correlators
\begar
 \chi(p) &=& \int e^{i p x} \< 0| T \Phi^{\dagger}(\frac{x}{2})
                        \Phi(-\frac{x}{2})|P_n\>, \\
 \bar{\chi}(p)&=& \int e^{-i p x} \<P_n| T \Phi(\frac{x}{2})
                        \Phi^{\dagger}(-\frac{x}{2})|0>.
\ea
Using the integral representation for the step function which is
included in the time ordered product, one derives eq. \pref{chisbar}.

Taking the equation for the Green function
\begar \plabel{bsf}
 iG_0 = -D_0 + D_0 K_0 G_0, 
\ea
with
\beg
 D_0 = \frac{ (2\pi)^4 \d^4(p-p') 
}{[(\frac{P_0}{2}+p_0)^2-E_p^2+i\e][(-\frac{P_0}{2}+p_0)^2-E_p^2+i\e]},
\ee
instead of that for the BS wave function and using again \pref{k0s} 
we find

\begar \plabel{Gscal}
G_0 = - F(p) \frac{G_C(\widehat{E},\p,\p\,')}{4 m}  F(p')
\ea
with
\beg
 \widehat{E} = \frac{P_0^2-4m^2}{4m}
\ee
and
\begar
 F(p) &=& \frac{\sqrt{E_p} (P_0^2-4E_p^2)}
  {[(\frac{P_0}{2}+p_0)^2-E_p^2+i\e][(-\frac{P_0}{2}+p_0)^2-E_p^2+i\e]}.
\ea
$G_C$ denotes the well known Coulomb Green function in momentum space.
These solutions can be used for a systematic BS perturbation theory for
scalar constituents, as will be demonstrated in the next section.

\subsection{Unstable Particles}
\plabel{unstpar}

As has been shown recently by the author \pcite{brig} for the fermionic 
case, an important simplification can be achieved in some calculations if the 
width of the bound state is already included in the zero order equation. 
Furthermore, if the width becomes comparable to the level shifts, this 
approach even becomes 
indispensible. For the scalar case this can be done by the replacement
\beg \plabel{root}
 E_p \to \sqrt{E_p^2-i \G m}.
\ee
While \pref{root} leads to expressions for the BS wave functions which contain
unpleasant expressions for the particle poles it has the advantage that
the propagator has the form as expected from the phase space of an
unstable particle. Furthermore the above calculation remains essentially
unchanged if we define the square root in \pref{root} to be that with
the negative imaginary part (clearly we demand $\G >0$ and $m>0$). 
Only the energy in the resulting equation for the Green function 
and thus in \pref{Gscal} changes to
\beg
  \widehat{E} = \frac{P_0^2-4m^2}{4m}+i\G. 
\ee
The eigenvalues for $P_0$ are
\beg \plabel{P0scal}
 P_{0,n} = 2 m \sqrt{1-\s_n^2-i \frac{\G}{m}} 
\approx 2m - m \s^2 -\frac{m \s_n^4}{4} + \frac{\G^2}{4m} - i \G
- i \frac{ \s_n^2 \G}{2} 
\ee

In the case of the fermions we managed to construct wave functions
independent of $\G$. This was possible because the small components
of the propagator containing $P_0-i\G$ instead of $P_0+i\G$ were projected
away by the choice of an appropriate kernel $K$.
This cannot be achieved in the scalar case and thus, surprisingly 
enough, the scalar wave functions look more complicated than the fermionic
ones. A version for a zero order equation for decaying particles 
where the propagator is chosen in close analogy to the fermionic case 
has been developed in \pcite{diss}. In our present work we, instead, 
proceed in the spirit of our generalized approach.

\section{Perturbation Theory}
\plabel{BSpert}

Perturbation theory for the BS equation starts
from the BR equation for the Green function $G_0$ (eq. \pref{bsf})
of the scattering of two fermions \pcite{BR2} which is
exactly solvable.
$D_0$ is the product of two zero order propagators, $K_0$ the corresponding kernel.
The exact Green function $G$ may be represented as
\beg \plabel{Reihe}
 G =\sum_{l} \chi_{nl}^{BS} \frac{1}{P_0 - P_n} \bar{\chi}_{nl}^{BS} + G_{reg}=G_0 \sum_{\n =0}^{\infty} (H G_0)^{\n} ,
\ee
where the corrections are contained in the insertions $H$ and $G_{reg}$ is
the part of $G$ regular at $P_0=P_n$.
It is easy to show that
$H$ can be expressed by the full kernel $K$ and the full propagators $D$:
\beg \plabel{H}
 H = -K + K_0 +iD^{-1}-iD_0^{-1}.
\ee
Thus the perturbation kernel is essentially the negative difference of 
the exact BS-kernel and of the zero order approximation.

Expanding both sides of equation \pref{Reihe} in powers of $P_0-P_n$,
the mass shift is obtained \pcite{Kum1,Lep77}:
\beg \plabel{dM}
 \D M - i\frac{\D \G}{2}= \< h_0 \> (1+\< h_1 \> ) + \< h_0 g_1 h_0 \> + O(h^3) .
\ee
Here the BS-expectation values are defined as e.g.
\begar
\< \<h\> \> &\equiv& \int \frac{d^4p}{(2\pi)^4} \int \frac{d^4p'}{(2\pi)^4}
           \bar{\chi}_{ij}(p) h_{ii'jj'}(p,p') \chi_{i'j'}(p'), \plabel{erww}
\ea
We emphasize the four-dimensional p-integrations which correspond to the generic
case, rather than the usual three dimensional ones in a completely
nonrelativistic expansion. We distinguish these two case by introducing
the notation $\< \< ...\> \> $ for a four-dimensional expectation
value and $\< ... \> $ for the usual nonrelativistic expectation
value
\beg
  \< V(\p,\p\,')  \> = \int \frac{d^3p}{(2 \pi)^3} \int \frac{d^3p'}{(2 \pi)^3} 
                        \phi^*(\p') V(\p, \p\,') \phi(\p)
\ee 
Of course, \pref{erww} reduces to an
ordinary "expectation value" involving $d^3p$ and $\Phi(\p)$, whenever
$h$ does not depend on $p_0$ and $p_0'$.

In \pref{dM} $h_i$ and $g_i$ represent the expansion coefficients of
 $H$ and $G_0$ near the pole at $P_n$, respectively, i.e.
\begar
 H&=& \sum_{m=0}^{\infty} h_m (P_0-P_n)^m \\
 G_0&=& \sum_{m=0}^{\infty} g_m (P_0-P_n)^{m-1} \plabel{gent}
\ea
Similar corrections arise for the wave functions \pcite{Kum1,Lep77}:
\beg  \plabel{chi1}
\chi^{(1)} = ( g_1 h_0 + \2  \< h_1 \> ) \chi^{(0)}
\ee

\subsection{Fine structure}

As an application of this perturbation theory as well as of the new zero
order equation for scalar particles developed in the last section, we will
present here the calculation of the fine structure of two stable scalar
particles interacting via a vector particle. Existing calculations
\pcite{Owen} rely on a mix of Fouldy-Wouthuysen transformation and the
iterated Salpeter perturbation theory. Our present approach is much more
transparent and allows in principle the inclusion of any higher order
effect in a straightforward manner. First we will calculate the fine
structure for two scalars of equal mass interacting by an abelian vector
field. Then we consider also the nonabelian case which could be of 
interest for the stop-antistop system.
In this case we will calculate the spectrum up to order $\a_s^4$.

\begin{figure}
\begin{center}
\leavevmode
\epsfxsize=13cm
\epsfbox{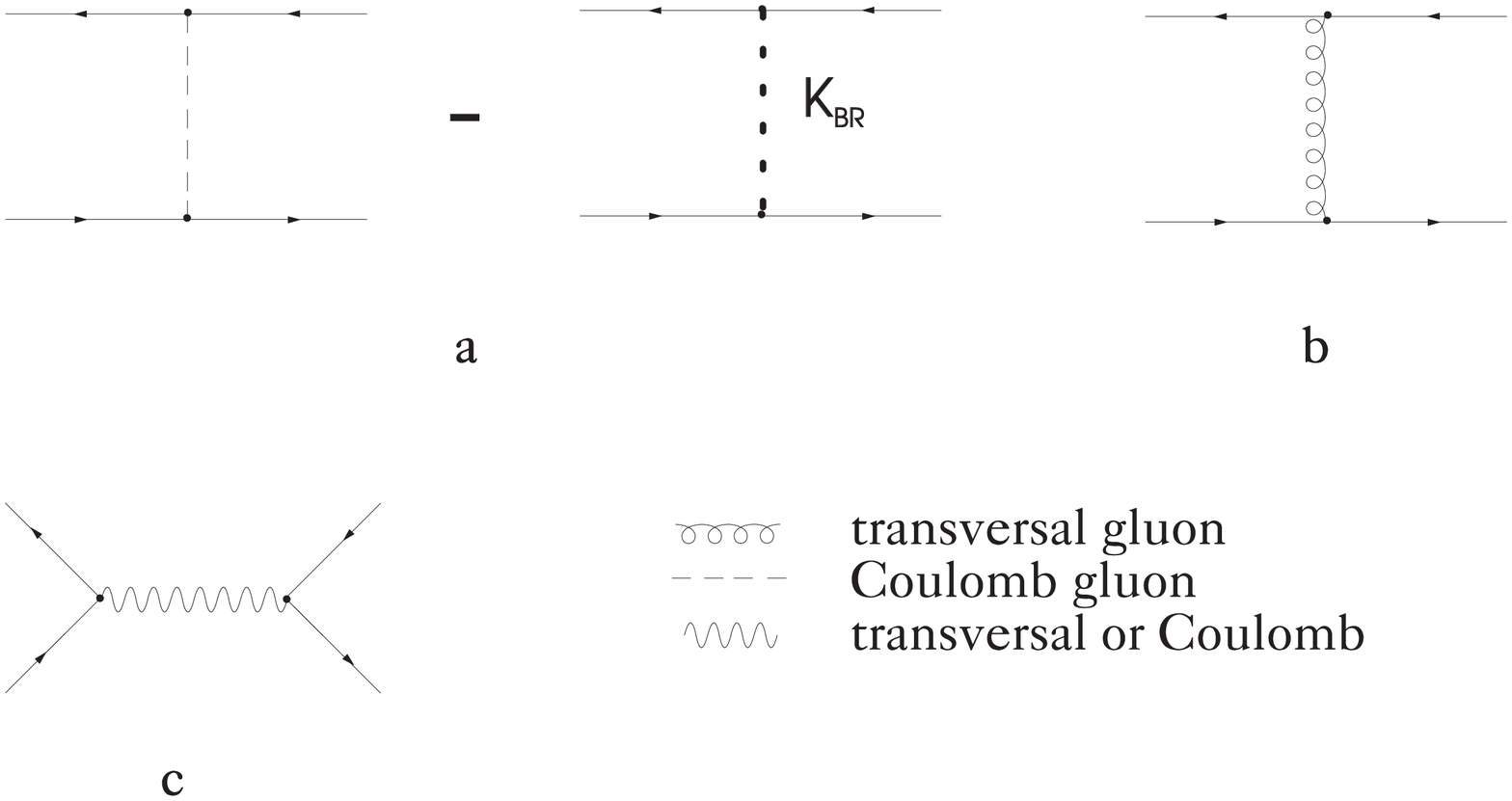}
\\
\centerline{ Fig. \refstepcounter{figure} \label{dfig2} \thefigure}
\end{center}
\end{figure}

Since in the zero order equation we have replaced the exact one Coulomb
exchange \pref{kcs} by $K_0$ as given in \pref{k0s} we have now to
calculate the contribution of $-K_C+K_0$ to the energy levels. This is
shown in fig. \pref{dfig2}a .
With
\begar
\<\<-K_C\>\> &=& - 4\pi \a \int \frac{d^4p}{(2\pi)^4} \frac{d^4p'}{(2\pi)^4}
                \bar{\chi}(p)  \frac{(P_0+p_0+p_0')
            (P_0-p_0-p_0')}{(\p-\p\,')^2} \chi(p') = \nn \\
         &=& -\< \frac{P_0^2+2E_p^2+2E_{p'}^2}{4 P_0 \sqrt{E_p E_{p'}}}
             \frac{4 \pi \a}{\q^2} \> =  \plabel{p0int} \\
         &=& -\< \left( \frac{2m}{P_0} - \frac{\s_n^2}{2} \right)
             \frac{4 \pi \a}{\q^2} \> \nn \\
\<\<K_0\>\> &=& \frac{2m}{P_0} \< \frac{4 \pi \a}{\q^2} \>
\ea
we obtain
\begar \plabel{relko}
 \D M_{C} &:=& \<\<-K_C+K_{0}\>\> = \frac{\s_n^2}{2} \< \frac{4 \pi
               \a}{\q^2} \> = \frac{m \a^4}{16 n^4}.
\ea
The fact that the p-integrations are well behaved and the result
is of $O(\a^4)$ proves the usefulness of our zero order kernel.

\begin{figure}
\begin{center}
\leavevmode
\epsfxsize=13cm
\epsfbox{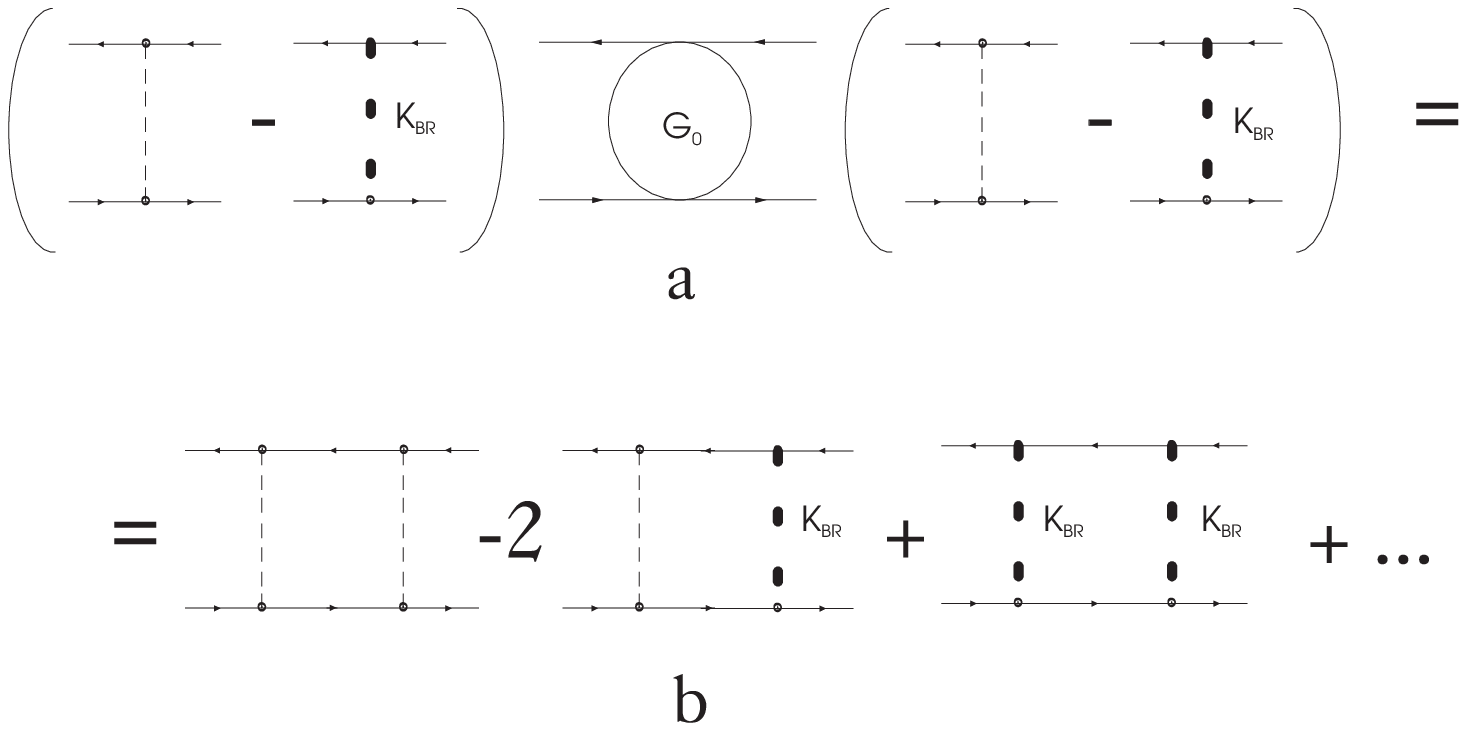}
\\
\centerline{ Fig. \refstepcounter{figure} \label{coulsec} \thefigure}
\end{center}
\end{figure}

The transverse gluon of fig. \ref{dfig2}b gives rise to a kernel
\beg
H_T = \frac{4 \pi \a}{q^2} \left( (\p+\p\,')^2 - \frac{(\p^2-\p\,'^2)^2}{\q^2}
      \right) 
\ee
Performing the zero component integrations exactly and expanding in terms
of the spatial momenta one obtains to leading order (c.f. \pcite{Land})
\begar
 \D M_T = \<\< H_T \>\> &=& - \frac{4 \pi \a}{m^2} \<
                   \frac{\p\,^2}{\q\,^2}-\frac{(\p \q)^2}{\q\,^4} \>\\
&=& m \a^4 \left(\frac{1}{8 n^4} + \frac{\d_{l0}}{8 n^3} - \frac{3}{16 n^3
(l+\2)} \right).
\ea
Due to the fact that scalars can only form spin zero bound states,
the the annihilation graph into one gauge particle (with spin one) contributes
only for p-waves and thus is supressed by two additional powers in $\a$.
Furthermore, as in the fermionic case, it vanishes for the nonabelian 
theory due to the color trace since the bound states are color singlets. 
As can be seen from the above
results the contribution of the transverse gauge field is equal for 
fermions and bosons. However, the relativistic correction to the Coulomb 
exchange appears to be different. Let us therfore check the contribution
of this Coulomb correction from second order perturbation theory
(fig. \ref{coulsec}.a). These contributions give only rise to $O(\a^5 \ln \a)$ 
effects in the fermionic theory.
Since the leading Coulomb singularity is cancelled 
we may hope that we can replace the Green function by the free 
propagator. Indeed it can be shown that the next terms of the 
Green function give only higher order
contributions.

\begin{figure}
\begin{center}
\leavevmode
\epsfxsize=13cm
\epsfbox{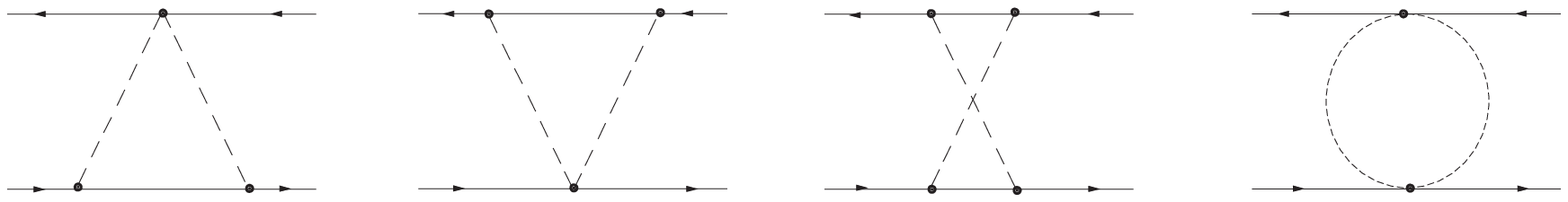}
\\
\centerline{ Fig. \refstepcounter{figure} \label{box} \thefigure}
\end{center}
\end{figure}

Due to the presence of the zero component momentum in the scalar-Coulomb
gluon vertex we observer that the contribution from fig.\ref{coulsec}.b diverges linearly.
However, it is an easy exercise to show that in the sum of graphs
(fig.\ref{coulsec}.b + fig. \ref{box}) this linear divergence cancels.
Thus we regularise all the single graphs, sum up and find a finite
result. 
We have used dimensional regularization as well as a one dimensional
Pauli-Villars regularization. Both give the same result for the 
finte parts of the intgrals.

\begar
 \D M_{box}  &=& \<\< h_0^{(3)} \>\> + \< \< (-K_C+K_0) g_1 (-K_C+K_0) \> \> 
\ea

\begar
 \<\< h_0 \>\> = \<\< -i \int \frac{d^3k}{(2 \pi)^3}
                \frac{I_0}{\vec{k}\,^2 (\q -\vec{k})^2} \>\>  
 \plabel{Ibox}
\ea
where $I_0$ is decomposed according to fig. \ref{box} for a generic
$SU(N)$ theory:

\begar
I_0^{(\ref{box}.a)} &=& \frac{C_F}{2 N} \int_{k_0} 
 \frac{(P+2 p' +k_0)(-P+2 p'+k_0)(P+p+p'-k_0)
        (P+2 p-k_0)}{[(\frac{P}{2}+ p-k)^2-m^2][(-\frac{P}{2}+ p'+ k)^2-m^2]} = \nn \\
 &\approx&  \frac{C_F}{2 N} ( \frac{\L}{2} - 2 i m) \\
I_0^{(\ref{box}.b)} &=& (C_F^2-\frac{C_F}{2 N}) \int_{k_0}
\left(  \frac{(P+p+p'-k_0)(P+2 p-k_0)}{[(\frac{P}{2}+p-k)^2-m^2]}\right.+ \nn \\
 & & \qquad + \left. \frac{ (-P+2 p' +k_0)(-P +p + p'+k_0)}{[(-\frac{P}{2}+ p'+k)^2-m^2]} 
\right) = \nn \\
 &\approx&  2 (C_F^2-\frac{C_F}{2 N})  ( \frac{\L}{2} - i m) \\
I_0^{(\ref{box}.c)} &=& -(C_F^2-\frac{C_F}{2 N})\int_{k_0}  = \nn \\
 &=&  -(C_F^2-\frac{C_F}{2 N}) \frac{\L}{2}  
\ea
Using the abbreviations
\begar
  \int_{k_0} &=& \int \frac{d k_0}{2\pi} \frac{\L^2}{k_0^2+\L^2} \plabel{PVreg} \\
  C_F &=& \frac{N^2-1}{2 N}  \plabel{CF}
\ea
we have written the result for Pauli Villars regularization to make 
the cancellation of the liner divergent parts obvious. 

For the double Coulomb exchange graph from fig. \ref{coulsec} we obtain
for the time component integral
\begar
I_0^{(\ref{coulsec}.b)} &=& -C_F^2 \int_{k_0} 
 \frac{[( p_{0}'+k_0)^2-P^2 + 4m \sqrt{E_k E_{p'}}][( p_{0}+k_0)^2-P^2 + 
      4m \sqrt{E_k E_p}]}{[(\frac{P}{2}+k_0)^2-E_k^2] 
         [(\frac{-P}{2}+k_0)^2-E_k^2]} \nn \\
 &=&  -C^2_F ( \frac{\L}{2} - i m) 
\ea

Collecting everything from above we have 
\begar
I_0 = I_0^{(\ref{box}.a)} + I_0^{(\ref{box}.b)} +I_0^{(\ref{box}.c)} +
        I_0^{(\ref{coulsec}.b)}   = - C_F^2 i m ,
\ea 
which leads with eq. \pref{Ibox},\pref{CF} and \pref{alph} immediately to the result
\begar
 \D M_{abelian box}  &=& - \frac{m \a^4}{16 n^3 (l+\2)}
\ea

The net result for the spectrum of two scalars bound by an abelian gauge
field is equal to that of ref \pcite{Owen}. However, we showed which 
graphs contribute in a pure
BS approach which can be used as a basis for any higher order calculation.
  
We have also checked the derivative $\6K_0/\6 P_0$ contributing to $h_1$ 
and the X-graphs of fig. (\ref{dfig3}.g) with transverse gauge particles
for possible contributions. Our estimates only 
yield contributions to higher order. Due to mass and wave function 
renormalization we can further assume that the graphs of fig. 
(\ref{dfig3}.e,f)
give only contributions to $O(\a^5 \ln \a)$ as in the fermionic case 
\pcite{Male}.
Possible large contributions of lighter particles to the vacuum 
polarization as depicted in (\ref{dfig3}.c) can be treated as in the 
fermionic case \pcite{KM1}.

In supersymmetric theories a $|\Phi|^4$ term is part of the lagrangian.
Clearly it can be put in by hand into the Langragian of an ordinary 
qantum field theory. 
The contribution from an interaction term of the form $-\l/2 
(\Phi^{\dagger}T^a\Phi)(\Phi^{\dagger}T^a\Phi)$ is easily calculated:
\beg
  \D M_X =- C_F \l  \frac{m \a^3}{32 \pi n^3} \d_{l0} 
\ee
and gives a contribution of the same form as the Darwin term (usually
interpreted as a zitterbewegung contribution) which is suppressed by
two orders in $\a$ in the scalar theory. There may exist a small 
chance that this term may be helpful for the determination of the 
supersymmety parameters of the theory contained in $\l$.

For an ordinary quantum field theory without a direct interaction
on the tree level it was shown first by Rohrlich \pcite{Rohr} that a
counter term of this form is needed for the scattering of two 
scalars (e.g the graphs of fig. \ref{box} and the first of fig. 
\ref{coulsec}.b  with Photons in Feynman gauge).
It is interesting to note that in Coulomb gauge the divergencies for 
the Coulomb photons cancel and the only divergent graph is the one 
of fig \ref{box}.c with transverse photons. 

The spectrum calculated so far is common for the abelian and the 
nonabelian theory. Collecting all pieces a we have 
\begar \plabel{resab}
\D M = \D M_{F,nl}^j +\frac{m \a^4}{8} \left( \frac{5}{4 n^4} + 
\frac{\d_{l0}}{n^3}(1-\frac{C_F \l}{4\pi \a}) -\frac{4}{n^3 (l+\2)}  \right)
\ea
where $\D M_{F,nl}^j$ originates in the contribution of $j$ light fermions 
to the vacuum polarization and can be found in \pcite{KM1}.

It has been pointed out first in \pcite{Duncan} that
in the case of a nonabelian gauge field further corrections may arise due
to the gluon splitting vertices. The $O(\a^3)$ corrections from fig. 
(\ref{dfig3}.a, b) as well as the $O(\a^4)$ corrections from the 
corresponding two loop graphs are obviously the same as in the 
fermionic case. The vertex correction shown below in fig.
(\ref{dfig3}.d) has been calculated in \pcite{Duncan} for the fermionic
case.

Here we will give a calculation of the same contribution for scalar
constituents. 
After performing the color trace the perturbation kernel for the second
graph in fig. (\ref{dfig3}.d) reads  
\beg
 H_{\ref{dfig3}.d,2} = -8ig^4 \int \frac{d^4 k}{(2 \pi)^4}
 \frac{(P_0+p_0+p_0'-k_0)(-P_0+p_0+p_0')}{\q\,^2 (\vec{k}-\q)^2
  [(\frac{P}{2}+p+k)^2-m^2] k^2} \left( -(\p \q) + \frac{(\p \vec{k})(\q
   \vec{k})}{\vec{k}\,^2} \right).
\ee
Performing the $k_0$ integration and using the scaling
\begar
 P_0 &\to& 2m + O(\a^2) \nn \\
 p_0 &\to&  \a^2 p_0  \nn  \\
 \vec{k} &\to& \a \vec{k} \nn
\ea
to extract the leading contribution in $\a$ we find that
\beg
  H_{\ref{dfig3}.d,2} = - \frac{g^2 m}{2} \frac{\p\q}{|\q|^3}.
\ee
Adding the similar contribution from the first graph in fig. (\ref{dfig3}.d)
gives
\beg
  H_{\ref{dfig3}.d} = - \frac{9\pi^2\a^2 m}{|\q|}.
\ee
This result differs by a factor $4m^2$ from the fermionic result which
is compensated by a corresponding difference in the wave functions to give
eventually precisely the same result as in the fermionic case
\beg
 \D M = \< \frac{9\pi^2\a^2}{4m|\q|} \> = \frac{9 m \a^4}{32 n ^3 (l+\2)} .
\ee
In view of the fact that the result 
depends only on the angular momentum and not on the spin this seems 
reasonable. However, we have seen in the case of the Darwin term
this kind of reasoning sometimes fails.

Proceeding to the graph of fig. (\ref{dfig3}.h) we observe that in 
contrast to the fermionic case the zero component integration
develops Coulomb divergencies like the abelian contributions.
Since this integrations are a little bit cumbersome in dimensional 
regularization we scetch the calculation in the appendix.
It turns out finally that the box graph contribution
in fig. (\ref{dfig3}.h) gives the same result as in the fermionc case,
which was calculated recently \pcite{hpap}.
\beg
\<\< H_{(\ref{dfig3}.h)} \>\> = -\frac{81}{128} \pi(12-\pi^2) 
            \< \frac{\a^3}{\q^2} \>
\ee
The box graph with two
Coulomb lines crossed, vanishes due to the color trace. Another box
graph with the Coulomb vertices on one scalar line replaced by a seagull
vertex can be shown to contribute to $O(\a^5)$. However, they are in 
principle needed to cancel the Coulomb singularities.   

Thus the difference in the spectrum of the scalar bound state to $O(\a^4)$ 
compared to the fermionic case is entirely due to the graphs also present in 
the abelian theory discussed above.

\begin{center}
\leavevmode
\epsfysize=12cm
\epsfbox{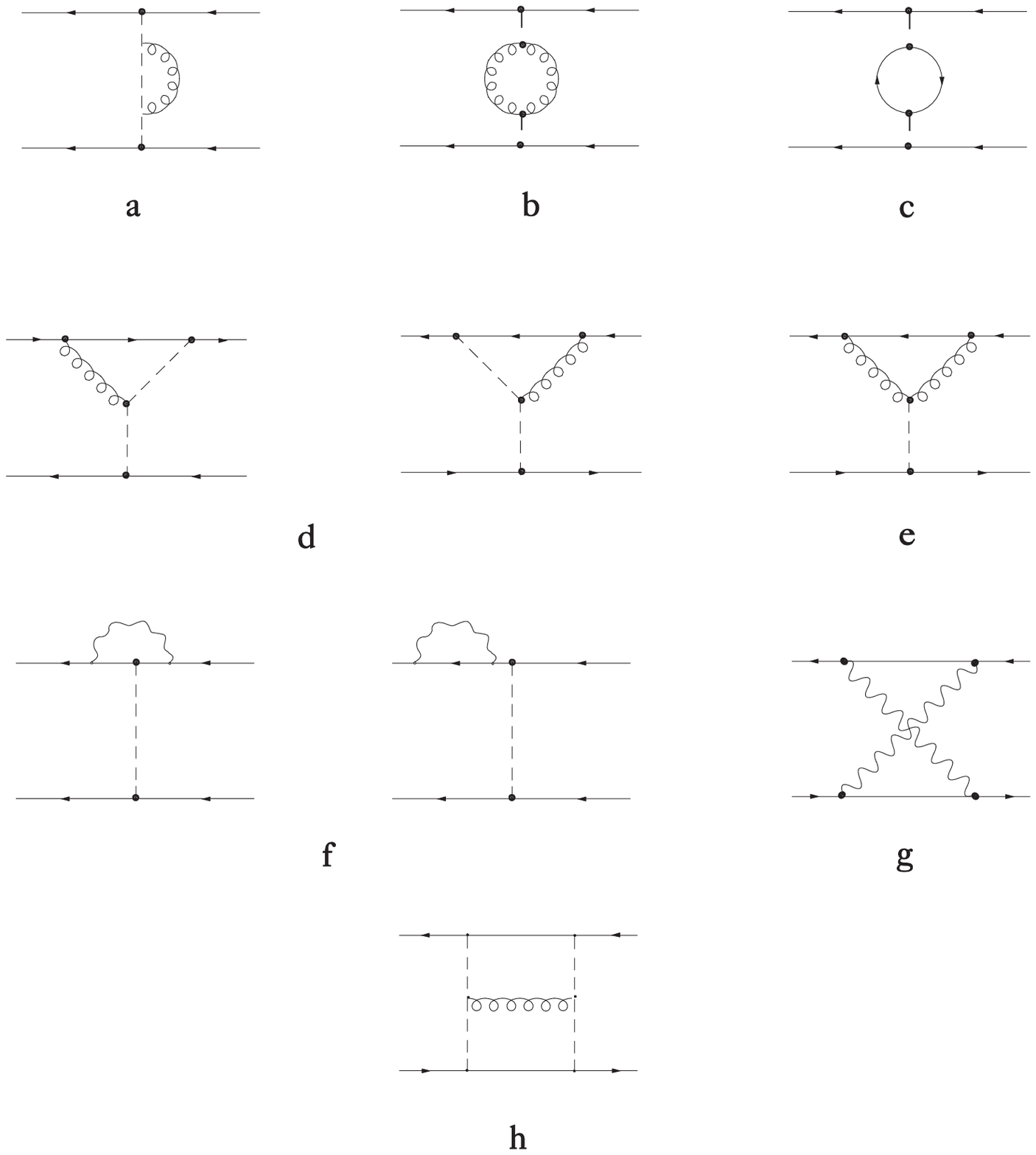}
\\
\centerline{ Fig. \refstepcounter{figure} \plabel{dfig3} \thefigure}
\end{center}

\section{Bound state corrections to the decay width}

Assuming that the scalar particle under consideration decays
into two other particles, the decay width is the imaginary part of the 
self energy function $\S$ at the mass shell. Focusing on the stop quark
a possible scenario could be $\tilde{t}_R \to b + \tilde{\chi}_i$ 
\pcite{Bartl}. We shall be interested in terms of the order $O(\a^2 \G)$
where $\G$ is the tree level decay width.
The first part of 
the perturbation kernel due to the exact inverse propagator 
$p^2-m^2-\S(p^2)$ for the bound state corrections to the decay 
width reads 
\begar
 H_1 &=& iD^{-1}-iD_0^{-1} = \nn \\
   &\approx& -2i (2\pi)^4 \d^4(p_1-p_2)  \S'(m^2) (p_1^2-m^2)(p_2^2-m^2) \plabel{Hbskorr}
\ea
In the derivation of eq. \pref{Hbskorr} we expanded the self energy 
function around the mass shell
\beg
 \S(p^2) = \S(m^2) + \S'(m^2) (p^2-m^2) + O( (p^2-m^2)^2)
\ee
and we assumed that the decay width used in the zero order equation
(e.g. in $D_0$) is given by
\beg
 \G = -\Im \frac{\S(m^2)}{m}.
\ee

As has been first shown in \pcite{topdec}, the gauge dependence contained
in the off shell contribution $\S'$ is cancelled by parts of the vertex 
correction  depicted in fig. \ref{dfig3}.f.
It gives rise to a perturbation kernel
\beg
 H_2 = \L_0 \frac{4\pi\a}{\q\,^2} (-P_0+p_0+p_0'),
\ee
with $\L_0$ representing the vertex correction. The color trace is 
already included in $\a$.
As in the fermionic case \pcite{brig} it is possible to derive a Ward 
identity which guaranties the cancellation of the gauge dependent terms
(the $T^a$ 's are the $SU(N)$ generators)
\begar
 \L_{\m}^a(p,q=0)  =- 2 g T^a p_{\m} \frac{\6}{\6 p^2} \S(p^2), \nn \\
 \Im \L_0(p=(m,\vec{0}),q=0) = - 2m \Im \S'(m^2).
\ea
But the detailed calculation shows differences to the fermionic case:
The sum of the contributions from
$H_1$ and $H_2$ vanishes to the desired order with the help of the zero 
order equation
\beg
 \Im \<\< H_1 + H_2 \>\> \approx 0.
\ee
On the other hand, we observed above that the wave functions for
decaying fermions and scalars were very different. While it was possible to obtain
the same wave functions for decaying fermions and for stable ones, in
the bosonic case we used wave functions explicitely containing the decay 
width (cf. sect. \ref{unstpar}). Thus we have to reexamine the 
relativistic corrections to the energy levels. Among the contributions 
considered in the last section only the relativistic Coulomb correction
fig. (\ref{dfig2}.a) can produce corrections to $O(\a^2\G)$.

It is easy to see that the only difference comes from the fact that the
perturbation has to be taken at the position of the pole \pref{P0scal}
which leads to the replacement
\beg
 \s_n^2 \to \s_n^2 + i \frac{\G}{m}.
\ee
in eq. \pref{p0int}.
We thus get a relativistic correction to the decay width of the bound 
state
\beg \plabel{erg}
 \D \G_{\ref{dfig2}.a} = -\frac{\G \a^2}{2 n^2}
\ee
This has to be added to the $O(\a^2 \G)$ term of \pref{P0scal} to
yield the final result for the boundstate correction to the decay width:
\beg 
 \D \G = -\frac{\G \a^2}{4 n^2}
\ee

This result generalizes the result of  \pcite{brig,topdec} to the
bosonic case. We can thus say that the effect of the bound state
corrections to the decay width can be interpreted entirely as a time
dilatation effect as was first conjectured for the fermionic theory
\pcite{Jeza}.

\section{Conclusion}

We have presented a consistent formalism for the calculation of bound 
state properties for scalar particles interaction with an abelian or
nonabelian spin one vector field. This is done by deriving a solvable
relativistic zero order equation similar to that of Barbieri and Remiddi
both for stabel and unstable scalars.
Based on this equation a systematic perturbation theoy can be built
which allows especially the calculation of the position of the bound 
state poles to higher orders. 

Using this approach the bound state spectrum was calculated to $O(\a^4)$.
We found that we had to take into account
the abelian box graphs to this order. This is not the case in the 
fermionic theory. All the relativistic Coulomb corrections only 
reproduce the $\p\,^4$ term from the expansion of $\sqrt{m^2+\p\,^2}$
indicating that a fully relativistic formulation is not really economic 
for the lowest orders in perturbation theory. However, the advantage of 
the presented formalism is that it is straigtformward applicable
to any higher order calculation. We calculated also the nonabelian 
contributions to $O(\a^4)$. 
Furthermore our approach makes possible the calculation of the bound state 
corrections to the decay width of weakly decaying scalar particles.
We show that - as in the fermionic case - the inlcusion of a finite,
constant decay width in the zero order equation simplifies the 
problem of the bound state correction to the decay width in a profound way. 
It is now possible to clearly isolate the underlying cancellation 
mechanism which automatically gives a gauge independent result 
which can be interpreted as time dilatation
alone. We can thus gereralize the theorem on the bound state
corrections for the decay width to the scalar case: The leading 
bound state corrections for weakly bound systems of unstable scalars (with decays like 
$\tilde{t}_R \to b + \tilde{\chi}_i$) are {\it always} of the form \pref{erg}.  

It would be very interesting to observe a particle where the above 
mentioned predictions could be tested. Today it seems that the 
stop-antistop system could be a candidate. It will be heavy enough to 
allow a perturbative treatment even for the nonabelian case. Whether
the decay width will be small enough to allow a detailed study of the
spectrum remains open to speculation at present. But even for a quite 
large decay width the scalar-scalar potential will provide the basis for
interesting threshold calculations for this case \pcite{WM1}, 
analogous to the ones for the top-antitop system \pcite{Jeza,ttth}.

\vspace{.5cm}

{\bf Acknowledgement}: I would like to thank Prof. W. Kummer for helpful 
discussions and a careful reading of the manuscript. Furthermore I am 
grateful to D. Raunikar and L. Widhalm for pointing out to me an error
in the calculation of the abelian box graphs and for checking large
parts of the calculations.

\begin{appendix}
\section{Zero component integrations for the nonabelian box graph}

The diagram \ref{dfig3}.h leads to the energy component integrals
\begar
I_0 := \int \frac{dk_0}{2\pi} \int \frac{dt_0}{2\pi} \frac{
(P_0+2p_0-t_0)(P_0+2p_0-q_0-t_0)(-P_0+2p_0-q_0-k_0)(-P_0+2p_0-k_0)}
{[(P_0/2+p_0-t_0)^2-E_{\p-\vec{t}}^2][(-P_0/2+p_0-k_0)^2-E_{\p-\vec{k}}^2]
 [(t_0-k_0)^2-(\vec{t}-\vec{k})^2]} \nn \\
\plabel{anga}
\ea
This integral is power counting logarithmic divergent, but it turns out 
that the first integration is finite which leads to a linear divergent 
second integration.
After scaling eq. \pref{anga} reduces to 
\begar
  I_0 = -\int \frac{dk_0}{2\pi} \int \frac{dt_0}{2\pi} 
  \frac{(2m-t_0)(2m+k_0)}{(t_0-i\e)(k_0+i\e) 
[(t_0-k_0)^2-(\vec{t}-\vec{k})^2+i\e]}
\ea
To make this integral accessible for the methods of dimensional 
regularisation we use the following trick 
\begar 
  \int \frac{dt_0}{2\pi} \frac{1}{t_0 \pm i\e} = \lim_{\m \to 0} \int 
\frac{dt_0}{2\pi} \frac{t_0 \pm \m}{t_0-\m^2+i\e}.
\ea
Performing first the $t_0$ integration we have
\begar
I_0     &=& I_{0,1}+I_{0,2} \nn \\
I_{0,1} &=& \int \frac{dk_0}{2\pi} \frac{(2m+k_0)(k_0+\m}{k_0^2-\m^2+i\e} 
I_{t_0} \nn \\
I_{0,2} &=& \int \frac{dk_0}{2\pi} I_{t_0}  \nn \\
I_{t_0} &=& -\frac{2mi \G(2-\frac{D}{2})}{(4\pi)^{\frac{D}{2}}}
            \int_0^1 dx \frac{(xk_0-\m)[x(1-x)]^{\frac{D}{2}-2}}
             {[-k_0^2 + \frac{\m^2}{x} + \frac{|\vec{t}-\vec{k}|^2}{1-x}]^{
              2-\frac{D}{2}} } - \frac{i \G(1-\frac{D}{2})}{(4\pi)^{\frac{D}{2}}}
             |\vec{t}-\vec{k}|^{D-2} \nn
\ea   
The limit $\m \to 0$ has to be performed very carefully to obtain
\begar
I_{0,1} &=& \frac{2m^2}{|\vec{t}-\vec{k}|^2}-\frac{m}{2|\vec{t}-\vec{k}|}  \nn \\
I_{0,2} &=& -\frac{m}{2|\vec{t}-\vec{k}|}
\ea
Thus we have to the desired accuracy
\beg
 I_0 = \frac{2m^2}{|\vec{t}-\vec{k}|^2}.
\ee
It should be noted that dimensional regularization does not show up 
linear divergencies. 
Instead the use of the regularization \pref{PVreg} leads to a 
visible liner divergent term (of higher order in $\a$) 
$i \L/(8 |\vec{t}-\vec{k}|)$ which has to be cancelled by similar 
contributions from graphs where the two Coulomb gloun vertices on 
one or both scalar lines are double Coulomb vertices.

\end{appendix}

\end{document}